\documentclass{article}
\usepackage[preprint]{spconf}
\usepackage{amsmath,graphicx}
\toappear{\begin{minipage}{\textwidth}\footnotesize \copyright2021 IEEE.  Personal use of this material is permitted.  Permission from IEEE must be obtained for all other uses, in any current or future media, including reprinting/republishing this material for advertising or promotional purposes, creating new collective works, for resale or redistribution to servers or lists, or reuse of any copyrighted component of this work in other works.\end{minipage}}

\usepackage[table]{xcolor}
\usepackage{tikz}
\usetikzlibrary{positioning}
\usetikzlibrary{calc}
\usetikzlibrary{arrows} 
\usetikzlibrary{arrows.meta}
\usetikzlibrary{decorations.markings}
\usepackage{booktabs}
\usepackage{tikzscale} 
\usepackage{floatrow}
\usepackage{amssymb}
\usepackage[hyphens]{url}
\usepackage{hyperref}
\newfloatcommand{capbtabbox}{table}[][\FBwidth]
\newfloatcommand{capffigbox}{figure}[][\FBwidth]

\newcolumntype{C}[1]{@{}>{\columncolor{white}[0pt]\centering\arraybackslash}p{#1}@{}}

\title{Lightweight and interpretable neural modeling of an audio distortion effect using hyperconditioned differentiable biquads}
\name{Shahan Nercessian, Andy Sarroff, and Kurt James Werner}
\address{iZotope, Inc., Cambridge, MA}

\begin{document}
\maketitle
\begin{abstract}
In this work, we propose using differentiable cascaded biquads to model an audio distortion effect.  We extend trainable infinite impulse response (IIR) filters to the hyperconditioned case, in which a transformation is learned to directly map external parameters of the distortion effect to its internal filter and gain parameters, along with activations necessary to ensure filter stability.  We propose a novel, efficient training scheme of IIR filters by means of a Fourier transform.  Our models have significantly fewer parameters and reduced complexity relative to more traditional black-box neural audio effect modeling methodologies using finite impulse response filters.  Our smallest, best-performing model adequately models a BOSS MT-2 pedal at 44.1 kHz, using a total of 40 biquads and only 210 parameters.  Its model parameters are interpretable, can be related back to the original analog audio circuit, and can even be intuitively altered by machine learning non-specialists after model training.  Quantitative and qualitative results illustrate the effectiveness of the proposed method.
\end{abstract}
\section{Introduction}
\label{sec:intro}
Digital emulation of analog audio circuits is a topic of extensive study within the audio signal processing community.  Such emulations allow for the widespread use of audio effects without the need for dedicated hardware.  Virtual analog techniques including wave digital filters \cite{Fettweis1986,Werner2016,Werner2018} are often used to model said circuits.  While these approaches can provide high fidelity emulations, they require significant domain knowledge, and often employ expensive online solvers \cite{Vesa1}.

Deep learning has emerged as an alternative, data-driven technique for modeling audio effects. In \cite{Martinez1, Martinez2}, generic black-box architectures were proposed to model a variety of effects, but use bidirectional layers having poor real-time implications.  A feedforward variant of WaveNet was proposed for emulating a tube amp \cite{Vesa1} and various distortion pedals \cite{Vesa2}.  The architecture boasts real-time performance and can model external parameters of the underlying effects, but still imposes a significant amount of complexity. Moreover, these black-box models lack interpretability, making them difficult to understand or troubleshoot.

Recently, there has been a push towards implementing differentiable audio processors in deep learning frameworks, enabling end-to-end training \cite{DDSP}. \cite{NI} considered learning fixed infinite impulse response (IIR) filters to approximate a single configuration of an audio effect, and thus, cannot model external parameters.  We have also found (and verified through talks with the original authors) that their implementation of IIR filters as recurrent layers restricts the number of filters that can be practically learned, limiting their applicability \cite{personalCommunication}.

In this paper, we propose modeling an audio distortion effect using multiple differentiable biquads.  We continue the trend towards lightweight, interpretable audio deep learning models, using several IIR filters as building blocks.  Unlike previous attempts to learn trainable IIR filters using deep learning \cite{NI, Bavu}, we use hyperconditioning (HC) \cite{HCNAF, Blow} in order to capture changes in filtering stages based on the user-facing parameters controlling the effect.  We consider many cascaded biquad (CB) representations and introduce activation functions necessary for their stable training.  We propose an efficient training scheme using frequency domain sampling, which is more tractable than a na{\"i}ve, prohibitively slow training using recurrent layers. The resulting models are lightweight and interpretable, yet perform competitively with a black-box baseline model.

The remainder of this paper reads as follows: We review biquads in Sec.\ \ref{sec:background}. We introduce the proposed method in Sec.\ \ref{sec:model}. We illustrate its effectiveness, comparing it to a black-box baseline in Sec.\ \ref{sec:results}. We draw conclusions in Sec.\ \ref{sec:conclusions}.

\section{Background Information}
\label{sec:background}
\begin{figure}[tb]
    \centering
    \includegraphics{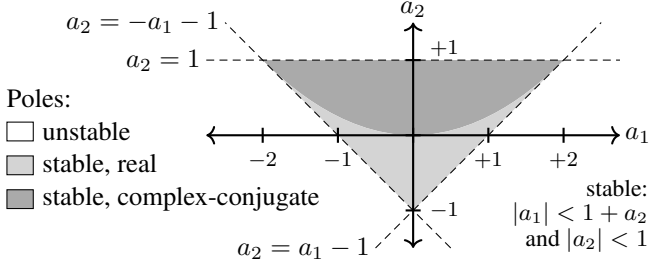}
    \caption{Biquad triangle ensuring filter stability.}
    \label{fig:biquadTriangle}
\end{figure}

Biquads ($2$nd order IIR filters) have the difference equation
\begin{align}
\begin{split}
y[n] = {b_0}x[n]+{b_1}x[n-1]+{b_2}x[n-2] \\ - {a_1}y[n-1]-{a_2}y[n-2]
\end{split}
\end{align}
where coefficients $ \mathbf{b} = [b_0, b_1, b_2] $ and $ \mathbf{a} = [1, a_1, a_2] $ are the feedforward and feedback gains of the filter, respectively \cite{Mitra}.  For real $\mathbf{b}$ and $\mathbf{a}$, the transfer function of a biquad is given by
\begin{align}
\label{eqn:transfer}
H(z) = \frac{b_0+b_1z^{-1}+b_2z^{-2}}{1+a_1z^{-1}+a_2z^{-2}} = b_0\frac{(z-q_0)(z-q_1)}{(z-p_0)(z-p_1)}
\end{align}
where pairs of zeros $\{q_0, q_1\}$ and poles $\{p_0, p_1\}$ dictate the nature of the filtering, and $b_0$ controls its gain.  Stability is ensured so long as both $|p_0|<1$  and $|p_1|<1$, achieved when $a_1$ and $a_2$ are constrained to be inside the so-called biquad triangle (Fig.\ \ref{fig:biquadTriangle}).

The frequency response of a biquad can be evaluated over a vector of linearly-spaced digital frequencies $\mathbf{\omega}$ by evaluating Eq.\ \ref{eqn:transfer} with $ e^{j\mathbf{\omega}}$ as its argument.  Higher order IIR filters can be created by cascading $K$ biquads in series.  The composite complex frequency response of such a cascade is given by
\begin{align}
\label{eqn:freqz}
{{H_\text{cascade}(e^{j\mathbf{\omega}})} ={ \textstyle \prod_{k=0}^{K-1} H_{k}(e^{j\mathbf{\omega}})}}
\end{align}
which we will refer to as the cascaded \texttt{freqz} operation.  While biquads model arbitrary $2$nd order filters, the RBJ cookbook \cite{RBJ} defines explicit formulae for parametric peaking and shelving filters with specified center/cutoff frequency $f$ in Hz, gain $g$ in decibels (dB), and Q/slope $Q$ at a given sampling rate $f_\text{SR}$.  The parameterization of these filters is interpretable for end-users and stable by design.  The design of cascaded peaking and shelving filters has been explored in several works \cite{Nercessian, Udo, Abel}, which form the basis for the ubiquitous parametric equalizer (EQ).

\section{Modeling using differentiable biquads}
\label{sec:model}
\begin{figure}[tb]
    \centering
    \!\!\!\!\!\includegraphics{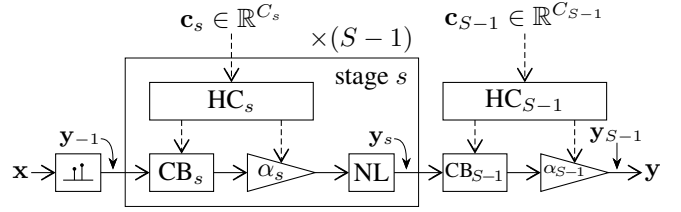}
    \caption{Proposed neural network model architecture.}
    \label{fig:architecture}
\end{figure}

A high-level block diagram of the proposed model is illustrated in Fig.\ \ref{fig:architecture}.  The input audio $\mathbf{x}$ is subjected to a parametric delay layer, yielding $\mathbf{y}_{-1}$, followed by $S$ filtering stages.  Each filtering stage $s$ consists of a cascade of $K$ biquads, a multiplicative gain $\alpha_s$ (specified in dB), and a $\mathrm{tanh}$ nonlinearity (NL), except for the final stage where the NL is omitted.  Each biquad is described by $P$ filter parameters, whose value depends on the specific representation that is used.  In addition to the audio signal from the previous stage $\mathbf{y}_{s-1}$, each filtering stage also takes as input $C_s$ external parameters from a user, forming the conditioning vector $\mathbf{c}_s$, which acts as a signal capable of altering the underlying parameters of the filtering stage.  The vector $\mathbf{c}_s$ can be made to vary over time to accommodate parameter automation.  The output of the final stage $\mathbf{{y}}=\mathbf{y}_{S-1}$ is an estimate of the ground truth output from the effect that is being modeled.

\subsection{Parametric delay layer}
The input is passed through a delay layer parameterized by 2 trainable parameters: a delay (in non-integer samples) and a linear gain.  The delay layer is critical for accommodating any potential timing offsets between inputs and outputs in the dataset.  The delay parameter maps to a linear fractional delay filter consisting of 2 taps which sum to 1 \cite{PASP}.  The delay filter computation could be extended to a more generic $\mathrm{sinc}$ interpolation kernel or be replaced with a non-sparse convolution, but neither are considered here in order to minimize complexity.  The filtered result is multiplied by a linear gain term, which in particular, gives the network the ability to potentially invert the signal.  As the $\mathrm{tanh}$ NL used in each block is symmetric, we found this to be sufficient for modeling a circuit containing an odd number of inverting amplifier stages.

\subsection{Hyperconditioned blocks}
Previous attempts to learn IIR filters in audio deep learning \cite{NI, Bavu} have only learned fixed filters which do not change with external user input.  Conventionally, conditioning information is injected into networks via addition \cite{WaveNet}, concatenation \cite{AutoVC}, or adaptive instance normalization \cite{AdaIN}.  Conversely, HC learns transformations of conditioning signals which ultimately control the underlying model weights (for instance, the coefficients of a convolutional kernel) \cite{HCNAF}. This type of adaptation is more direct and expressive, validating its use in recent audio style transfer applications \cite{Blow}.  We inject conditioning information (i.e.\ the values of user-facing effect parameters scaled to be in the range $[0, 1]$) into the model by applying HC to the cascade of biquads and gain in a filtering stage.  Specifically, we transform conditioning vectors $\mathbf{c}_s$ to underlying parameters at filtering stage $s$ using a single affine transformation, implemented as a $1\times1$ convolution.

Inspection of the schematic of the circuit being modeled can help appropriately set the value of $S$ and $K$, and to determine which filtering stages have external parameters as input.  Injecting the knowledge of the conditioning sites into the model can further reduce parameter counts and impose more inductive bias into the model.  For blocks with $C_s > 0$, we use HC networks to transform conditioning signals into their corresponding parameters.  Otherwise, we simply use trainable weights which learn a fixed gain and filtering for the stage.

\subsection{Cascaded biquad representations}
We consider 3 parameterizations of CBs in this work. While \cite{NI} suggests that filter stability can be largely ensured with proper initialization when learning fixed IIR filter configurations  (as network optimization will naturally promote learning the stable solution), this is not sufficient for hyperconditioned filters that change with user input, as in this work.  We ensure stable training and inference by subjecting parameters to representation-specific activations imposing their stability constraints.  We generically refer to these activations as \texttt{p2c}, characterizing all of the operations that are performed to map parameters to their ultimate filter coefficients and gains.  Small ``epsilon" terms avoid marginal stability and/or undefined behavior, but we omit them here for readability.

\subsubsection{Coefficient representation}
In the $1$st parameterization, we characterize the $k$th biquad at stage $s$ by their coefficient vectors $\mathbf{b}_{s, k} $ and $ \mathbf{a}_{s, k} $.  Since each CB is followed by a gain, we set $b_{0,s,k}=1$ for all filters with no remarkable loss to model expression.  The vector of filter parameters is $\mathbf{v}_{s,k} = [b_{1,s,k}, b_{2,s,k}, a_{1,s,k}, a_{2,s,k}] $ and $P=4$.
In order to maintain stability, we must enforce that $a_{1,s,k}$ and $a_{2,s,k}$ are inside the biquad triangle.  This is accomplished by performing the following activations in order:
\begin{align}
\label{eqn:triangle}
a_{1,s,k} \longleftarrow &2\cdot\tanh{a_{1,s,k}} \\
a_{2,s,k} \longleftarrow & \left[\left(2-| a_{1,s,k}|\right)\tanh{a_{2,s,k}} + |a_{1,s,k}|\right]/2
\end{align}

\subsubsection{Pole/zero representation}
The $2$nd parameterization characterizes each biquad by the real and imaginary part of a single pole and zero, constraining the remaining pole and zero to be their complex conjugates. In this case, $\mathbf{v}_{s,k} = [\Re\{q_{s,k}\}, \Im\{q_{s,k}\}, \Re\{p_{s,k}\}, \Im\{p_{s,k}\}] $, and again, $P=4$. Stability is ensured by constraining $|p_{s, k}| < 1$ using the activation function
\begin{align}
\label{eqn:norm}
p_{s,k} \longleftarrow p_{s,k}\tanh{\left(|p_{s,k}|\right)}/|p_{s,k}|
\end{align}
and we compute $\mathbf{b}_{s, k} = [1, -2\Re\{q_{s,k}\}, \Re\{q_{s,k}\}^2+\Im\{q_{s,k}\}^2]$, $\mathbf{a}_{s, k} = [1, -2\Re\{p_{s,k}\}, \Re\{p_{s,k}\}^2+\Im\{p_{s,k}\}^2]$.

\begin{figure*}
    \begin{floatrow}
        \capbtabbox{%
            \centering
            \small
            \begin{tabular}{lll}
                \toprule
                $s$ & $C_s$ & Parameter name \\ \midrule
                2 & 1 & \texttt{DIST} \\
                6 & 2 & \texttt{LOW}, \texttt{HIGH} \\
                7 & 2 & \texttt{MID}, \texttt{MID FREQ} \\
                8 & 1 & \texttt{LEVEL} \\
                \bottomrule
            \end{tabular}
        }{%
          \caption{Conditioning sites.}
            \label{tab:cond}%
        }\!\!\!%
        \capbtabbox{%
            \centering
            \small
            \begin{tabular}{lll}
                \toprule
                Model & Params. & MSE \\ \midrule
                Coefficient & 274 & 0.1708 \\
                Pole/zero & 274 & 0.0885 \\
                Param.\ EQ & \cellcolor{black!20}{\bf{210}} & \cellcolor{black!10}\bf{0.0629}\\
                WaveNet & 22960 & \cellcolor{black!20}{\bf{0.0088}} \\
                \bottomrule
            \end{tabular}
        }{%
          \caption{Model comparisons.}
            \label{tab:models}
        }\ \ \ %
        \capffigbox{%
            \includegraphics{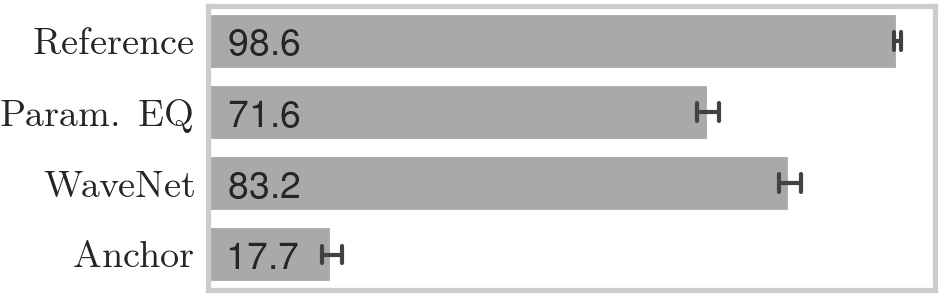}
        }{%
          \caption{MUSHRA scores with 95\% confidence intervals.}%
          \label{fig:mushra}
        }%
    \end{floatrow}
\end{figure*}

\subsubsection{Parametric EQ representation}
In our final parameterization, we model each CB with a parametric EQ.  Each parametric EQ consists of a low shelf, a high shelf, and $K-2$ peaking filters.
In this case, we have $\mathbf{v}_{s,k} = [f_{s,k}, g_{s,k}, Q_{s,k}] $ and $P=3$.  The parametric EQ parameters are converted to biquad filter coefficients using the RBJ cookbook formulae \cite{RBJ}.

We apply an activation on $f_{s,k}$ such that it is in $[0, f_\text{SR}/2]$ and $f_{s,0} \leq f_{s,1} \leq ... \leq f_{s,K-1}$, defined as
\newcommand{\round}[1]{\ensuremath{\lfloor#1\rceil}}
\begin{align}
\label{eqn:freq}
f_{s,k} \longleftarrow (f_\text{SR}/K){\textstyle \sum_{k=0}^{K-1}}\big\lvert\round{f_{s,k}} - f_{s,k}\big\rvert
\end{align}
where $\round{\cdot}$ denotes the round operator.  No activation is applied to $g_{s,k}$, though gains specified in dB are inherently converted back to a linear scale in the RBJ cookbook formulae.  Values for $Q_{s,k}$ must be non-negative, and for shelving filters, must also be $\leq1$.  Accordingly, $Q_{s,k}$ is subjected to the activation 
\begin{align}
\label{eqn:qfactor}
Q_{s,k} \longleftarrow Q_{\text{max}, k}\cdot\sigma(Q_{s,k})
\end{align}
where $\sigma(\xi)=1/(1+\mathrm{e}^{-\xi})$ and
\begin{equation}
    Q_{\text{max}, k}= 
\begin{cases}
    3, & \text{if } 0 < k < K-1\\
    1,              & \text{otherwise}
\end{cases}\ .
\end{equation}
This parameterization has one fewer degree of freedom than the other representations but is more interpretable.

\subsection{Implementation and training}
It is becoming increasingly well-known that differentiable IIR filters can be constructed exactly using recurrent layers \cite{NI, Bavu, Nercessian}.  If we consider the number of filters which we may want to cascade, and that such layers would operate on audio samples at $f_\text{SR} \geq 44.1$ kHz, it is apparent that this would be prohibitively slow to train in practice.  Alternatively, we propose training differentiable biquads by implementing the cascaded \texttt{freqz} computation in Eq.\ \ref{eqn:freqz} and \texttt{p2c} procedures for each representation with differentiable operators.  Setting $\mathbf{\omega} $ to match the frequency axis of an underlying $N$-point fast Fourier transform (FFT) allows us to perform filtering in the frequency domain.  If $N$ is made large enough, the amount of time-aliasing incurred via frequency domain sampling of the system response is small enough that it sufficiently approximates the true IIR response (at least for training purposes). This methodology was initially considered in \cite{Nercessian} for the task of parametric EQ matching, and we have now extended it to the more general problem of (time-domain) audio effect modeling.  A short-time Fourier transform, to this end, allows us to train on arbitrary length signals.  During inference, filtering stage parameters are inferred from input parameter settings, and the resulting filters are implemented recursively.

\subsection{Model complexity and interpretability}
The number of parameters for the models proposed here is $2 + {\textstyle \sum_{s=0}^{S-1}} (1 + KP)(1+C_s)$.  While neural networks are often criticized for their lack of interpretability, our models are transparent and intuitive.  Particularly with the parametric EQ representation, each stage of the model can be hand-tuned by a machine learning non-specialist after training.  Note that the biases of HC affine transformations correspond to the ``quiescent" state of a stage with user inputs set to 0.  Therefore, by altering the biases of hyperconditioned filtering stages or the parameters of fixed ones, we can globally scale gains and resonances, and/or shift frequency responses to the left/right.  The ``action'' of external controls can be altered by adjusting the proper elements of the learned HC matrices.

\section{Experimental results}
\label{sec:results}
We seek to emulate the BOSS MT-2 distortion pedal, which has $S=10$ cascaded filtering stages and $C_{\text{max}}=6$ external parameters.  The parameters and their conditioning sites in the circuit are summarized in Tab.\ \ref{tab:cond}. We use an internally collected dataset of direct input guitar signals operating at $f_\text{SR}=44.1$ kHz, segmenting them into 1 s clips. Distorted audio is generated using a SPICE model of the analog circuit.  A large dataset is created by applying the effect with several random parameter settings to each clip.

In addition to the proposed models, we train a feedforward variant of WaveNet  \cite{Vesa2}, serving as a baseline black-box model.  The $1$st layer of the model is a $1\times1$ convolutional layer outputting 16 channels, followed by 10 dilated convolutional layers, with dilation rate increasing by a factor of 2 at each layer.  The number of residual, skip, and dilation channels per layer is 16.  Dilation channels are generated using causal convolutions with a kernel size of 3 and standard gated-filter activation \cite{WaveNet}. External parameter settings comprise a conditioning vector of length $C_{\text{max}}$, which are injected into the gate and filter of each layer by means of $1\times1$ convolutions and addition.  Residual and skip channels are generated using $1\times1$ convolutions. Skip channels are mixed using a $1\times1$ convolution with linear activation, yielding the final output.  We train models for 100 epochs with a batch size of 16 and ADAM optimizer.  We consider all representations discussed here with $K=4$, $N=4096$, and HC as in Tab.\ \ref{tab:cond}.  As in \cite{NI}, we use a time-domain mean squared error (MSE) loss, noting that pre-emphasis could improve perceptual relevance \cite{Vesa3}.

\begin{figure}
    \includegraphics{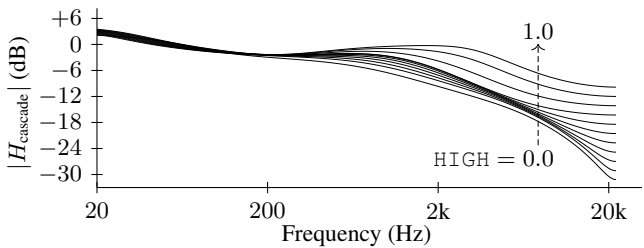}
    \caption{MT-2 magnitude responses ($s=6$), sweeping \texttt{HIGH}.}
        \label{fig:magnitudeReponses}
\end{figure}

We report the loss evaluated on a held-out test set alongside parameter counts for all models in Tab.\ \ref{tab:models}. While the baseline model does outperform the proposed models in terms of MSE, our models perform competitively given that they have on the order of 100 times fewer parameters and added interpretability.  Interestingly, we found that the parametric EQ representation actually outperformed the other representations mentioned here despite its lower parameter count.  It is also the most interpretable and user-tunable representation introduced here.  

We performed a MUltiple Stimuli with Hidden Reference and Anchor (MUSHRA) \cite{MUSHRA} listening test to evaluate models subjectively.  For each of 20 trials, participants were presented with test items and asked to rank how accurately each item sounded like the SPICE reference (0--100 scale).  We used the following test items for each trial: the SPICE reference itself, the input guitar signal, peak-normalized to $6.0$, and processed through a $\mathrm{tanh}$ NL (serving as an anchor), the output of the WaveNet baseline, and the output of our proposed method with parametric EQ representation.  We had a total of 28 participants involved in the music technology field.  Due to COVID-19 restrictions, participants conducted tests in their home environments with their own choice of headphones.

The results of our listening tests are shown in Fig.\ \ref{fig:mushra}.  We observe that all pairwise comparisons are statistically significant with 95\% confidence, and that our approach performs competitively, albeit just slightly worse than the baseline.  Informally, we observe that the emulation of the proposed method is still quite good, especially given its lower complexity, and that the WaveNet model often introduces unnatural resonances around 8 and 10 kHz.  Lastly, Fig.\ \ref{fig:magnitudeReponses} illustrates that HC can alter filter frequency responses in intuitive ways.  For example, increasing the \texttt{HIGH} parameter increases the response of stage 6 in the high end.  For audio examples and additional figures, please visit \href{https://sites.google.com/izotope.com/icassp2021-audio-demo}{\nolinkurl{https://sites.google.com/izotope.com/icassp2021-audio-demo}}.

\section{Conclusions}
\label{sec:conclusions}

We proposed modeling an audio effect using hyperconditioned differentiable biquads.  Several biquad representations were considered, and an FFT-based approximation enabled their efficient training.  Our models are lightweight and interpretable, and perform comparably to a black-box approach.  Future work will extend the use of our trainable IIR filtering methodologies to different audio modeling tasks.

\vfill
\pagebreak

\bibliographystyle{IEEEbib}
{
\bibliography{main}
}

\end{document}